\documentclass[pre,twocolumn,noshowpacs,amsmath,amssymb]{revtex4-2}
\usepackage{graphicx}
\usepackage{amsfonts}
\usepackage{dcolumn}
\usepackage{bm}
\usepackage{color}
\usepackage{hyperref}

\begin{document}
	
	\newcommand{\gin}[1]{{\bf\color{blue}#1}}
	\def\bc{\begin{center}}
		\def\ec{\end{center}}
	\def\bea{\begin{eqnarray}}
	\def\eea{\end{eqnarray}}
	\newcommand{\avg}[1]{\langle{#1}\rangle}
	\newcommand{\Avg}[1]{\left\langle{#1}\right\rangle}

\title{Entropy rate of random walks on complex networks under stochastic resetting}

\author{Yating Wang}
\author{Hanshuang Chen}\email{chenhshf@ahu.edu.cn}
\affiliation{School of Physics and Optoelectronic Engineering, Anhui University, Hefei 230601, China}

\begin{abstract}
Stochastic processes under resetting at random times have attracted a lot of attention in recent years and served as illustrations of nontrivial and interesting static and dynamic features of stochastic dynamics. In this paper, we aim to address how the entropy rate is affected by stochastic resetting in discrete-time Markovian processes, and explore nontrivial effects of the resetting in the mixing properties of a stochastic process. In particular, we consider resetting random walks on complex networks and compute the entropy rate as a function of the resetting probability. Interestingly, we find that the entropy rate can show a nonmonotonic dependence on the resetting probability. There exists an optimal resetting probability for which the entropy rate reaches a maximum. We also show that the maximum entropy rate can be larger than that of the maximal-entropy random walks on the same topology. Our study provides a new nontrivial effect of stochastic resetting on nonequilibrium statistical physics.   
\end{abstract}
\maketitle
\section{Introduction}
Stochastic resetting is ubiquitous in nature \cite{evans2020stochastic}. Maybe most of people have this experience: when one goes to work in the morning, (s)he searches for keys before going out. After a search without successes, it is likely to go back to the starting point of the search and try again. The motion of foraging animals can be also modeled as a resetting random walks \cite{RevModPhys.83.81,PhysRevLett.112.240601,JPA2022.55.274005,PhysRevLett.128.148301}. Indeed, animals tend to go back to some fixed location (e.g., to their nest) when searching for food. Other examples are realized in computer simulations in which random restarts are known to optimize search algorithms \cite{Luby1993,PhysRevLett.88.178701}, and in biology, e.g., to describe catastrophes in population dynamics \cite{Biophys.J.2010.98.1099}.

Since the seminal work by Evans and Majumdar \cite{evans2011diffusion}, stochastic resetting has received growing attention in the last decade (see \cite{evans2020stochastic,Gupta2022Review} for two recent reviews). A paradigmatic example in statistical physics is resetting Brownian motions where a diffusing particle is reset to its starting point at random times but with a constant rate. A finite resetting rate leads to a nonequilibrium stationary state with non-Gaussian fluctuations for the
particle position. The mean time to reach a given target for the first time can become finite and be minimized with respect to the resetting rate \cite{evans2011diffusion}. Reuveni first made a universal observation that the relative standard deviation associated with the first passage time of an optimally restarted process is always unity \cite{PhysRevLett.116.170601}. Pal and Reuveni
further showed a criterion under which restart has the ability to expedite the completion of a stochastic process \cite{pal2017first}. Interestingly, such a criterion can be
understood by so-called ``inspection paradox'' in probability theory \cite{JPA2022.55.021001}. Chechkin and Sokolov addressed random search via a renewal approach and showed that resetting is always beneficial if the probability of finding a target in absence of resetting decays slower than exponential \cite{chechkin2018random}. These nontrivial findings have triggered an enormous recent activities in the field, including searching  \cite{evans2011diffusion2,PhysRevLett.113.220602,ahmad2019first,NJP2016.18.033006,pal2016diffusion,PhysRevE.93.060102,PhysRevE.96.012126,evans2014diffusion,evans2018run,kumar2020active,PhysRevE.103.022135,de2020optimization,lauber2021first,JPA2021.54.404004,JPA2022.55.275002,PhysRevE.103.032107,arXiv:2202.04906,JPA2022.55.234001,PhysRevE.105.034109,huang2021random,PhysRevLett.128.200603}, fluctuating interfaces \cite{gupta2014fluctuating}, stochastic thermodynamics  \cite{fuchs2016stochastic,pal2017integral,gupta2020work}, chemical and biological
processes \cite{reuveni2014role,rotbart2015michaelis,PhysRevLett.128.148301}, large deviation \cite{meylahn2015large}, extremal statistics \cite{PhysRevE.103.052119,JStatMech2022.063202,JPA2022.55.034002} optimal control theory \cite{arXiv:2112.11416}, and single-particle experiments \cite{tal2020experimental,besga2020optimal}. 

Entropy is a fundamental concept in statistical physics. In the realm of complex networks, entropy has been introduced to measure the complexity of networks. The entropy of network ensembles quantifies the number of graphs with given structural features such as degree distribution, degree correlations, or community structure \cite{Bianconi_EPL2028,PhysRevE.79.036114,PhysRevE.80.045102,PhysRevE.82.011116}. The principle of maximum entropy has been used to construct exponential random graphs under different soft constraints \cite{PhysRevE.70.066117,NJP2011.13.083001,PhysRevLett.114.158701,PhysRevLett.115.268701,PhysRevE.87.062806,PhysRevE.93.062311,NatRevPhys2019.1.58}.  The entropy measures have been shown to be very useful for inference problems defined on networks \cite{PhysRevLett.120.198301}, and it has been successfully applied to the problem of assessing
the significance of features for network structure \cite{Bianconi_PNAS2009}.  
On the other hand, particular attentions have been paid to entropy rate of random walks on complex networks. Entropy rate is a measure to characterize the mixing properties of a stochastic process. In this context, an important issue arises how the entropy rate is maximized to design diffusion processes aiming at a well-mixed state. Burda \textit{et al.} proposed a maximum entropy random walks (MERW) in which all trajectories between two given points are equiprobable \cite{PhysRevLett.102.160602}. They showed that MERW indeed maximizes the entropy of trajectories, in contrast to standard
random walk (SRW), which has smaller entropy. The maximum entropy rate is precisely the topological entropy of the network \cite{Parry1964,Demetrius2005,PhysRevE.83.046117}. However, for MERW the walker needs to have a global knowledge of the network. G\'omez-Garde\~nes and Latora considered a degree-biased random walk and found that the entropy rate shows a unique maximum as a degree-biased parameter varies \cite{PhysRevE.78.065102}. Sinatra \textit{et al.} constructed the dynamics of random walks by solely using the degrees of first and second neighbors of the current node of the walker \cite{PhysRevE.83.030103}. They showed almost maximal-entropy random walks can indeed be obtained with a limited and local knowledge of the network. Zhao \textit{et al.} computed the entropy rate of various growing network models \cite{PhysRevE.84.066113}, and showed the
entropy rate changes its scaling with the system size when a growing network model has a phase transition. 

In the present work, we want to address a question about how stochastic resetting affects entropy rate of a Markov process and explore whether the resetting induces a novel effect on the entropy rate. We mainly focus on random walks on complex networks subject to stochastic resetting to a given node with a constant probability. We compute the entropy rate of the resetting random walks on diverse networks, including regular random networks, Cayley trees and degree heterogenous networks. We find that the entropy rate can be maximized at an intermediate level of resetting probability. In particular, the maximum entropy rate can be larger than that of MERW on the same network.

\section{Entropy rate of Markovian processes under resetting}
We consider discrete-time Markovian processes defined in a finite state space encoded by an $N \times N$ Markov matrix $\bm{W}$, whose entry $W_{ij}$ gives the transition probability from the $i$th state to the $j$th state. A trajectory $\omega_t$ denote a series of subsequent states that the system has visited in the past $t$ time step, i.e., $X_0 \to X_1 \to \dots \to X_{t-1} \to X_t$, where $X_i \in \left\{ {1,\cdots, N} \right\}  $ denote the state of the system at time $i$. The probability of the trajectory is given by

\begin{eqnarray}\label{eq1}
P\left[ {{\omega _t}} \right] = {P}\left( {{X_0}} \right)\prod\limits_{i = 1}^t {{W_{{X_{i - 1}}{X_i}}}} ,
\end{eqnarray}
where $P(X_0)$ denotes the probability that the system starts from the state $X_0$ at $t=0$, and $W_{{X_{i-1}}{X_i}}$ denotes the transition probability from the state $X_{i-1}$ to state $X_i$. The entropy is defined as
\begin{eqnarray}\label{eq2}
{H_t} =  - \sum\limits_{{\omega _t}} {P\left[ {{\omega _t}} \right]\ln P\left[ {{\omega _t}} \right]} ,
\end{eqnarray}
where the summation is over all possible trajectories of length $t$. Substituting Eq.\ref{eq1} into Eq.\ref{eq2}, we have 
\begin{eqnarray}\label{eq3}
{H_t} =&&  - \sum\limits_{{X_0}} {P}\left( {{X_0}} \right)\ln {P}\left( {{X_0}} \right)  \nonumber \\ && - t \sum\limits_{{X_0},{X_1}} {{P}\left( {{X_0}} \right){W_{{X_0}{X_1}}}\ln {W_{{X_0}{X_1}}}} ,
\end{eqnarray}
where we have utilized the properties of Markov matrix, $\sum\nolimits_j {{W_{ij}} = 1} $. For $t \gg 1$, the first term in Eq.\ref{eq3} can be ignored. At the same time, $P(X_0)$ can be substituted by the stationary distribution $P_s(X_0)$. Therefore, the entropy rate reads \cite{RevModPhys.85.1115} 
\begin{eqnarray}\label{eq4}
h = \mathop {\lim }\limits_{t \to \infty } \frac{{{H_t}}}{t} = -\sum\limits_{{X_0},{X_1}} {{P_s}\left( {{X_0}} \right){W_{{X_0}{X_1}}}\ln {W_{{X_0}{X_1}}}}.
\end{eqnarray}
It is known that the stationary distribution $P_s(X_0)$ is given by the normalized left eigenvector of the transition matrix $\bm{W}$ corresponding to the unit eigenvalue.

We now consider the Markovian processes in the presence of resetting. At each time step, the system either hops from one state to another according to the transition matrix $\bm{W}$ with a probability $1-\gamma$ or is reset to a given state $X_r$ with a complementary probability $\gamma$, in the sense that the transition probability from state $X_0$ to state $X_1$ is 
\begin{eqnarray}\label{eq4.1}
W^{R}_{X_0 X_1}=(1-\gamma) W_{X_0 X_1} + \gamma \delta_{X_1,X_r}.
\end{eqnarray}

Let us denote by $P_r(X_t,t|X_0)$ the probability that the system visits the state $X_t$ at time $t$ given that the system has started from the state $X_0$ at $t=0$. $P_r(X_t,t|X_0)$ can be connected to the occupation probability $P_0(X_t,t|X_0)$ without resetting via a first renewal equation \cite{pal2016diffusion,ahmad2019first,chechkin2018random,Chaos2021_31.093135,arXiv.2111.01330}, 
\begin{eqnarray}\label{eq5}
{P_r}\left( {{X_t},t|{X_0}} \right) &=& {\left( {1 - \gamma } \right)^t}{P_0}\left( {{X_t},t|{X_0}} \right) \nonumber \\ &+& \sum\limits_{t'=1}^{t} {{{\left( {1 - \gamma } \right)}^{t' - 1}}\gamma {P_r}\left( {{X_t},t - t'|{X_r}} \right)} .
\end{eqnarray}
The first term in Eq.\ref{eq5} accounts for the system is never reset up to time $t$ with the probability ${\left( {1 - \gamma } \right)^t}$, and the second
term accounts for the system is reset at time $t'$ for the first time with the probability ${{{\left( {1 - \gamma } \right)}^{t' - 1}}\gamma }$, after which the process starts anew from the resetting state for the remaining time $t-t'$. $P_0(X_t,t|X_0)$ is given by
\begin{eqnarray}\label{eq6}
{P_0}\left( {{X_t},t|{X_0}} \right) = {\left( {{\bm{W}^t}} \right)_{{X_0}{X_t}}}
\end{eqnarray}
Taking the Laplace transform for Eq.\ref{eq5}, $\tilde{f}(s)=\sum_{t=0}^{\infty} f(t) e^{-st}$, which yields
\begin{eqnarray}\label{eq7}
{{\tilde P}_r}\left( {{X_t},s|{X_0}} \right) &=& {{\tilde P}_0}\left( {{X_t},s'|{X_0}} \right)\nonumber \\ &+& \frac{{\gamma {e^{ - s}}}}{{1 - \left( {1 - \gamma } \right){e^{ - s}}}}{{\tilde P}_r}\left( {{X_t},s|{X_r}} \right),
\end{eqnarray}
where $s'=s-\ln(1-\gamma)$. Letting $X_0=X_r$ in Eq.\ref{eq7}, we have 
\begin{eqnarray}\label{eq8}
{{\tilde P}_r}\left( {{X_t},s|{X_r}} \right) = \frac{{1 - \left( {1 - \gamma } \right){e^{ - s}}}}{{1 - {e^{ - s}}}}{{\tilde P}_0}\left( {{X_t},s'|{X_r}} \right).
\end{eqnarray}
Substituting Eq.\ref{eq8} into Eq.\ref{eq7}, we obtain
\begin{eqnarray}\label{eq9}
{{\tilde P}_r}\left( {{X_t},s|{X_0}} \right) = {{\tilde P}_0}\left( {{X_t},s'|{X_0}} \right) + \frac{{\gamma {e^{ - s}}}}{{1 - {e^{ - s}}}}{{\tilde P}_0}\left( {{X_t},s'|{X_r}} \right). \nonumber \\
\end{eqnarray}
If the resetting state coincides with the initial state, $X_r=X_0$, Eq.\ref{eq9} simplifies to
\begin{eqnarray}\label{eq10}
{{\tilde P}_r}\left( {{X_t},s|{X_0}} \right) = \frac{{1 - \left( {1 - \gamma } \right){e^{ - s}}}}{{1 - {e^{ - s}}}}{{\tilde P}_0}\left( {{X_t},s'|{X_0}} \right),
\end{eqnarray}
where ${{\tilde P}_0}\left( {{X_t},s'|{X_0}} \right)$ can be calculated by Eq.\ref{eq6}, 
\begin{eqnarray}\label{eq11}
{{\tilde P}_0}\left( {{X_t},s'|{X_0}} \right) = \left[ {\bm{I} - \left( {1 - \gamma } \right){e^{ - s}}\bm{W}} \right]_{{X_0}{X_t}}^{ - 1}.
\end{eqnarray}
The stationary occupation probability can be obtain by taking the limit 
\begin{eqnarray}\label{eq12}
{P_s}\left( X \right) &=& \mathop {\lim }\limits_{s \to 0} \left( {1 - {e^{ - s}}} \right){{\tilde P}_r}\left( {X,s|{X_0}} \right) \nonumber \\& =& \gamma \left[ {\bm{I} - \left( {1 - \gamma } \right)\bm{W}} \right]_{{X_r}X}^{ - 1}.
\end{eqnarray}

Supposing that the transition matrix $\bm{W}$ can be eigen-decomposed, one has
\begin{eqnarray}\label{eq13}
\bm{W} =  \sum\limits_{\ell = 1}^N {\lambda_\ell \langle i | {{\phi _\ell}} \rangle } \langle {{{\bar \phi }_\ell}} | j\rangle ,
\end{eqnarray}
where $\lambda_\ell$ is the $\ell$th eigenvalue of the transition matrix $\bm{W}$, and the corresponding left eigenvector and right eigenvector are respectively $\langle {{\bar \phi }_\ell}|$ and $| {\phi_\ell}\rangle$, satisfying $\langle {{{\bar \phi }_\ell}} | {{\phi _m}} \rangle  = {\delta _{\ell m}}$ and $\sum_{\ell=1}^{N} |\phi_{\ell} \rangle \langle {\bar \phi}_{\ell}|=\bm{I}$. $\left| i \right\rangle $ denotes the canonical base with all its components equal to 0 except the $i$th one, which is equal to 1. Since $\bm{W}$ is a stochastic matrix satisfying the sum of each row of $\bm{W}$ equal to one, its maximal eigenvalue is equal to one. Without loss of generality, we let $\lambda_1=1$ and the values of other eigenvalues is less than one. The right eigenvector corresponding to $\lambda_1=1$ is simply given by $| {{\phi _1}} \rangle  = {\left( {1,1, \ldots ,1} \right)^\top}$, and the corresponding left eigenvector $\langle {\bar \phi _1} |$ gives the stationary occupation probability in the absence of resetting.

According to Eq.(\ref{eq13}), the stationary occupation probability in the presence of resetting is rewritten as 
\begin{eqnarray}\label{eq14}
P_s ( X) = \langle {{{\bar \phi }_1}} | X \rangle  + \gamma \sum\limits_{\ell = 2}^N {\frac{{\langle X_r | {{\phi _\ell}} \rangle \langle {{{\bar \phi }_\ell}} | X \rangle }}{{1 - \lambda _\ell^{}\left( {1 - \gamma } \right)}}} .
\end{eqnarray}
where the first term is the stationary occupation probability in the absence of resetting, and the second term is an nonequilibrium contribution due to the resetting processes. Finally, substituting Eq.(\ref{eq4.1}) and Eq.(\ref{eq14}) into Eq.(\ref{eq4}), we can compute the entropy rate for a resetting Markov process.

\section{Entropy rate of resetting random walks}
As a concrete example, we consider the resetting random walks (RRW) on an undirected and unweighted network. The dynamics is defined as follows \cite{PhysRevE.101.062147,PhysRevE.103.062126,Chaos2021_31.093135,JSM2022.053201}. At each time step, the walker either performs a standard random walk (SRW) between two neighboring nodes with a probability $1-\gamma$ or is reset to a given node $r$ with a complementary probability $\gamma$. For $\gamma=0$, the model recovers the SRW, where the transition matrix is written as $\bm{W}=\bm{D}^{-1}\bm{A}$, where $\bm{A}$ is the adjacency matrix of the underlying network, whose entries are defined as $A_{ij}=1$ if nodes $i$ and $j$ are connected and zero otherwise. $\bm{D}={\rm{diag}} \left\{ k_1,\cdots, k_N  \right\}$ is a diagonal matrix where $k_i=\sum_{j=1}^{N} A_{ij}$ is the degree of node $i$. For the SRW, it is known that $P_s(i)=k_{i}/(\langle k \rangle N)$ \cite{masuda2017random,PhysRevLett.92.118701,PhysRevE.87.012112}, where $\langle k \rangle$ is the average degree of the network. RRW on networks may has many applications in computer science and physics. Label propagation in machine learning algorithms \cite{Bautista2019}, or the famous PageRank \cite{Pagerank1998}, can be interpreted as a random walker with uniform resetting probability to all the nodes of the network. Based on hitting times under resetting, a recent study has made an application to network centrality \cite{avrachenkov2018hitting}.

\begin{figure}
	\centerline{\includegraphics*[width=1.0\columnwidth]{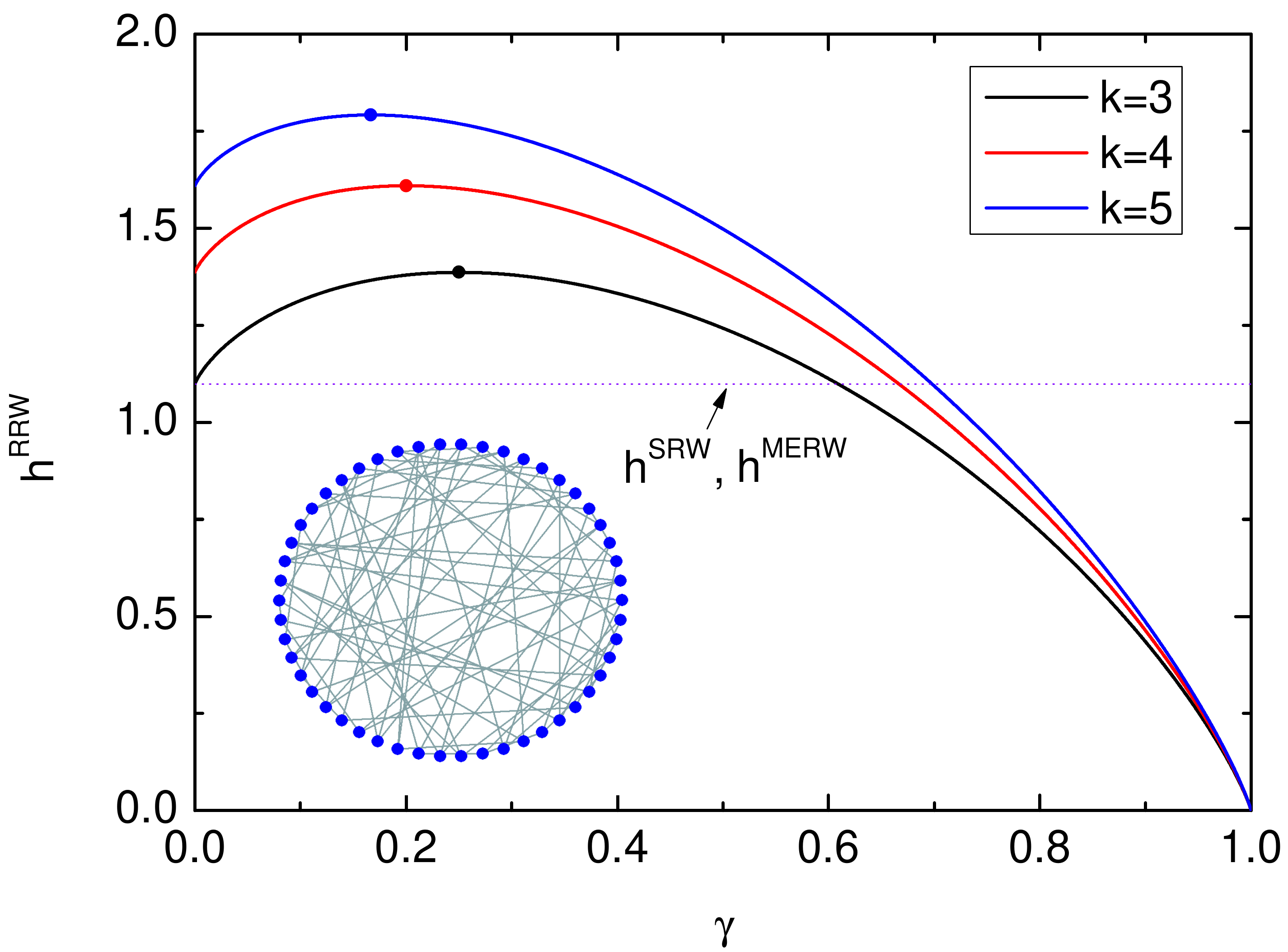}}
	\caption{Entropy rate $h^{RRW}$ of resetting random walks as a function of the resetting probability $\gamma$ on three different regular random networks in which each node is randomly connected to exactly $k$ neighbors. The inset shows an illustration of a regular random network with size $N=50$ and degree $k=3$. The symbols indicate the maximum entropy rate $h_{\max}^{RRW}=\ln(k+1)$ that occurs at  $\gamma_{{\rm{opt}}}=\frac{1}{k+1}$. The horizontal line indicates the value of entropy rate for the SRW or MERW with $k=3$, $h^{SRW}=h^{MERW}=\ln k$.  \label{fig1}}
\end{figure}

In terms of Eq.\ref{eq4}, one obtains the entropy rate of SRW,
\begin{eqnarray}\label{eq2.2}
{h^{SRW}} = \frac{{\langle {k\ln k} \rangle }}{{\langle k \rangle }}.
\end{eqnarray}
While for the RRW, one has
\begin{eqnarray}\label{eq2.3}
{h^{RRW}} =  - \sum\limits_{i=1}^N {{P_s}(i)\left( {1 - \gamma } \right)\ln \frac{{1 - \gamma }}{{{k_i}}} - \gamma \ln \gamma } .
\end{eqnarray}
To obtain $h^{RRW}$, one needs to compute the stationary occupation distribution $P_s(i)$ in terms of Eq.(\ref{eq14}), where the spectrum of the transition matrix $\bm{W}$ can be expressed in terms of the spectrum of the adjacency matrix $\bm{A}$. Letting $\Lambda_\ell$ be the $\ell$th eigenvalue of $\bm{A}$, and the associated eigenvector is $|\psi_\ell \rangle$, one has
\begin{eqnarray}
{\lambda _\ell} = \frac{{{\Lambda _\ell}}}{{{\Lambda _1}}}, \, \langle {i| {{\phi _\ell}} \rangle }  = \frac{{\langle {i| {{\psi _\ell}} \rangle } }}{{\langle {i| {{\psi _1}} \rangle } }}, \, \langle {{{\bar \phi }_\ell}} | j \rangle  = \langle {{\psi _\ell}} | j \rangle \langle {{\psi _1}} | 1 \rangle, 
\end{eqnarray}
for $\ell=1,\cdots,N$. $\Lambda _1 \geq \langle k \rangle$ is the largest eigenvalue of $\bm{A}$.

\begin{figure}
	\centerline{\includegraphics*[width=1.0\columnwidth]{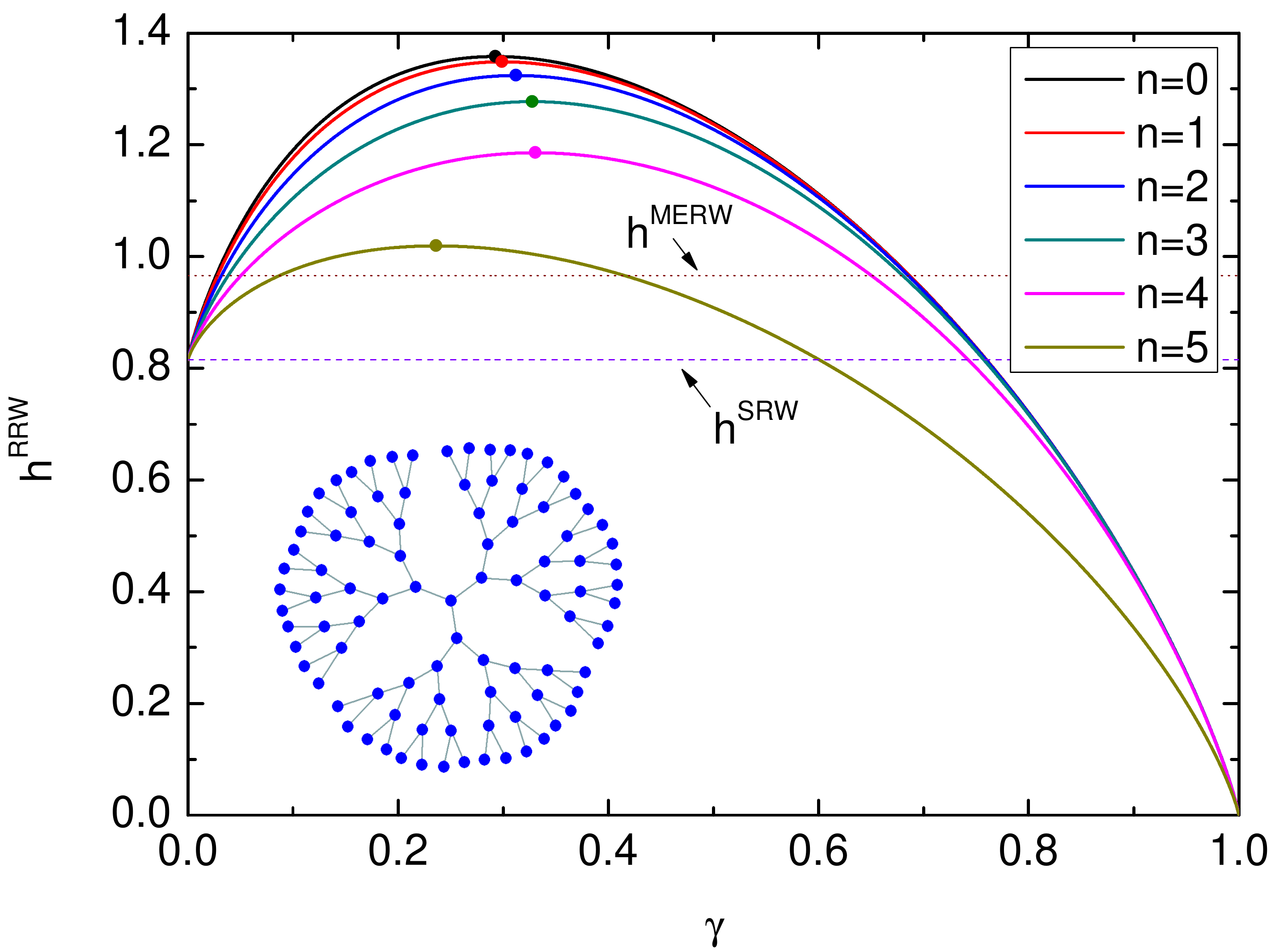}}
	\caption{Entropy rate $h^{RRW}$ of the resetting random walks as a function of the resetting probability $\gamma$ on a Cayley tree $C_{3,5}$ (see the inset). The symbols indicate the maximum entropy rate $h_{\max}^{RRW}$. The upper and lower horizontal lines indicate the values of entropy rate for the MERW and SRW, $h^{MERW}$ and $h^{SRW}$, respectively. Different lines represent that the only resetting node is placed at different shells: $n=0$ to $n=5$ from bottom to top. \label{fig2}}
\end{figure}

In particular, for regular random networks where each node is randomly connected to
exactly $k$ neighbors (see the inset of Fig.\ref{fig1} for an illustration), Eq.(\ref{eq2.3}) can be further simplified to
\begin{eqnarray}\label{eq2.4}
{h^{RRW}}  =  - \left( {1 - \gamma } \right)\ln \frac{{1 - \gamma }}{k} - \gamma \ln \gamma .
\end{eqnarray}
In Fig.\ref{fig1}, we show $h^{RRW}$ as a function of the resetting probability $\gamma$ on three different $k$-regular random networks: $k=3$, 4, and 5. The entropy rate defined in Eq.(\ref{eq2.4}) shows a nonmonotonic change with $\gamma$. For $\gamma \to 0$, $h^{RRW}$ recovers to the result in Eq.(\ref{eq2.2}) without resetting. In the opposite limit, $\gamma \to 1$, the system is always reset to a given node, such that the dynamics is deterministic and thus $h^{RRW} \to 0$.  There exists a maximum entropy rate at an intermediate value of $\gamma$, such that $h^{RRW}=h^{RRW}_{\max}$ at $\gamma=\gamma_{{\rm{opt}}}$. To obtain the maximum, we take the derivative of Eq.(\ref{eq2.4}) with respect to $\gamma$, and then let the derivative equal to zero. We obtain the maximum entropy rate $h_{\max}^{RRW}=\ln(k+1)$ that occurs at $\gamma_{{\rm{opt}}}=\frac{1}{k+1}$. We also perform Monte Carlo simulations for the resetting random walks and obtain the entropy rate for simulation date. We find that the simulation results are completely consistent with the theory. The simulation results are not shown in Fig.\ref{fig1} for the sake of clearness.  

Moreover, we compare the entropy rate of RRW with that of MERW \cite{PhysRevLett.102.160602}. For the MERW, the transition probability from node $i$ to node $j$ is defined as 
\begin{eqnarray}\label{eq2.5}
{W_{ij}^{MERW}} = \frac{{{A_{ij}}}}{\Lambda }\frac{{\langle {j| \psi_1  } \rangle }}{{\langle {i| \psi_1  } \rangle }},
\end{eqnarray}
where $\Lambda_1$ is the largest eigenvalue of the adjacency matrix $\bm{A}$, and $\langle i | \psi_1 \rangle$ is the $i$th component of eigenvector corresponding to $\Lambda_1$, as stated before. The MERW is biased in the sense that a walker follows a
link $(i,j )$ with a probability proportional to the importance of its ending node $j$, as measured by its eigenvector centrality $\langle {j|\psi_1} \rangle$. 
It is not hard to verify that the stationary distribution of MERW is $P_s(i)=\langle i | \psi_1 \rangle ^2$. Substituting this result and Eq.(\ref{eq2.5}) into Eq.(\ref{eq4}), one obtains the entropy rate of MERW, 
\begin{eqnarray}\label{eq2.6}
{h^{MERW}}  =  \ln \Lambda_1.
\end{eqnarray}

For regular random networks without degree fluctuation, $\Lambda_1=k$, and thus ${h^{MERW}}  =  \ln k$, coinciding with the entropy rate of SRW defined in Eq.(\ref{eq2.2}). Therefore, for entropy rate on $k$-regular random networks, we have
\begin{eqnarray}\label{eq2.7}
{h^{RRW}_{\max}}  = \ln(k+1) >{h^{MERW}} ={h^{SRW}}= \ln k.
\end{eqnarray}

\begin{figure*}
	\centerline{\includegraphics*[width=2.0\columnwidth]{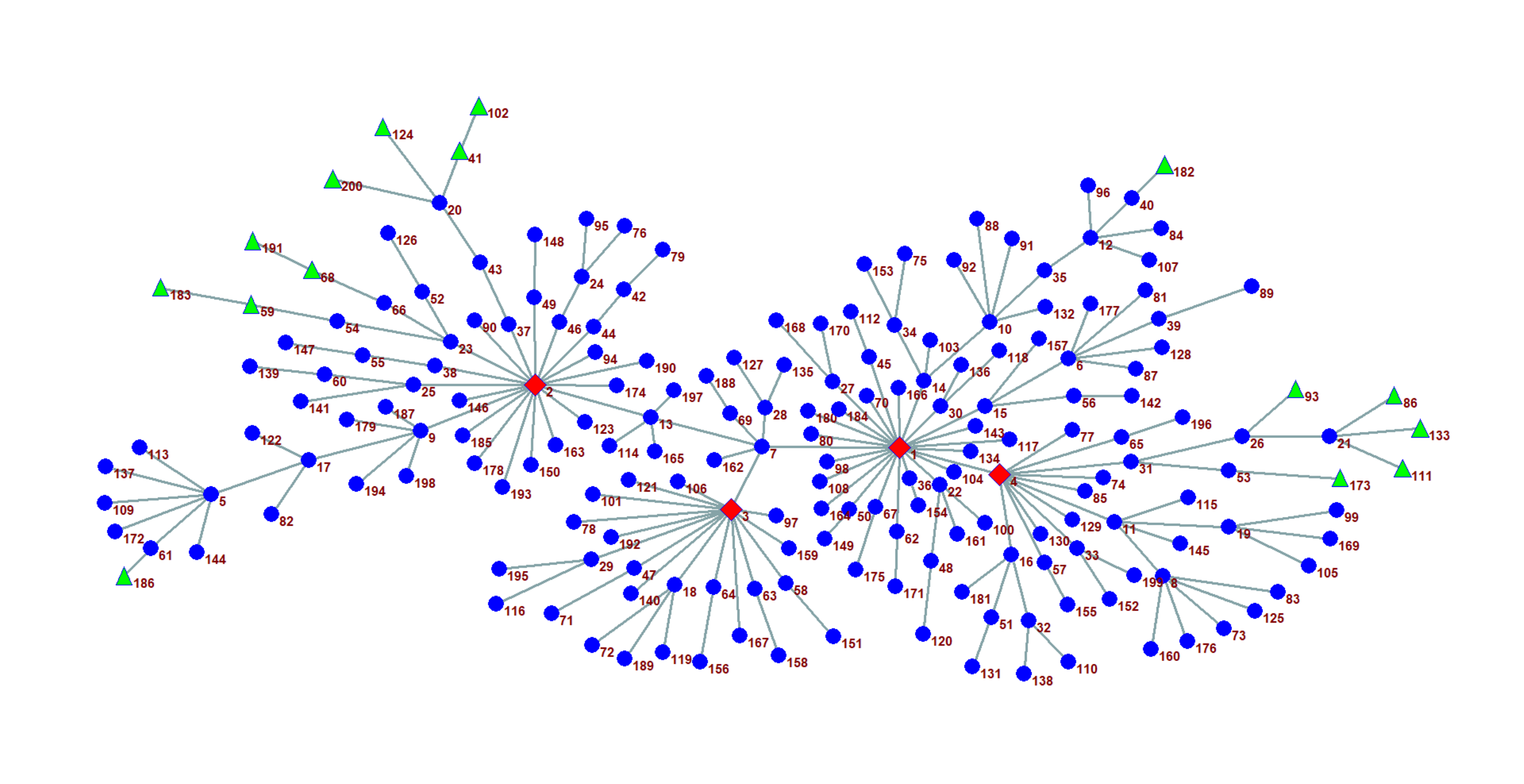}}
	\caption{A Barab\'asi-Albert (BA) network of size $N=200$ and average degree $\langle k \rangle =2$. The nodes are labelled in descending order by degrees of nodes. All nodes are classified into three types according to the entropy rate of resetting random walks by resetting to one of nodes. For four nodes with the largest dgrees (diamonds), the maximum entropy rate is larger then that of MERW,  $h_{\max}^{RRW}>h^{MERW}$. For 15 peripheral nodes (triangles), $h^{RRW}<h^{SRW}<h^{MERW}$ for any nonzero resetting probability. For the remaining nodes (circles), $h^{SRW}<h_{\max}^{RRW}<h^{MERW}$.      \label{fig3}}
\end{figure*}

On the other hand, one can see from Fig.\ref{fig1} that for $0<\gamma< \gamma_c$ the entropy rate of RRW on regular random networks is larger than that of SRW (or MERW). To determine $\gamma_c$, let $h^{RRW}=\ln k$, which yields a transcendental equation $k\gamma_c  = {\left( {1 - \gamma_c } \right)^{\frac{{\gamma_c  - 1}}{\gamma_c }}}$. For $k=4$, $\gamma_c=0.5$ is exact. For $k=3$ and $k=5$, numerically solving for the equation gives $\gamma_c \approx 0.609$ and 0.423, respectively. As $k$ increases, $\gamma_c$ decreases and approaches to zero as $k \to \infty$.

\begin{figure}
	\centerline{\includegraphics*[width=1.0\columnwidth]{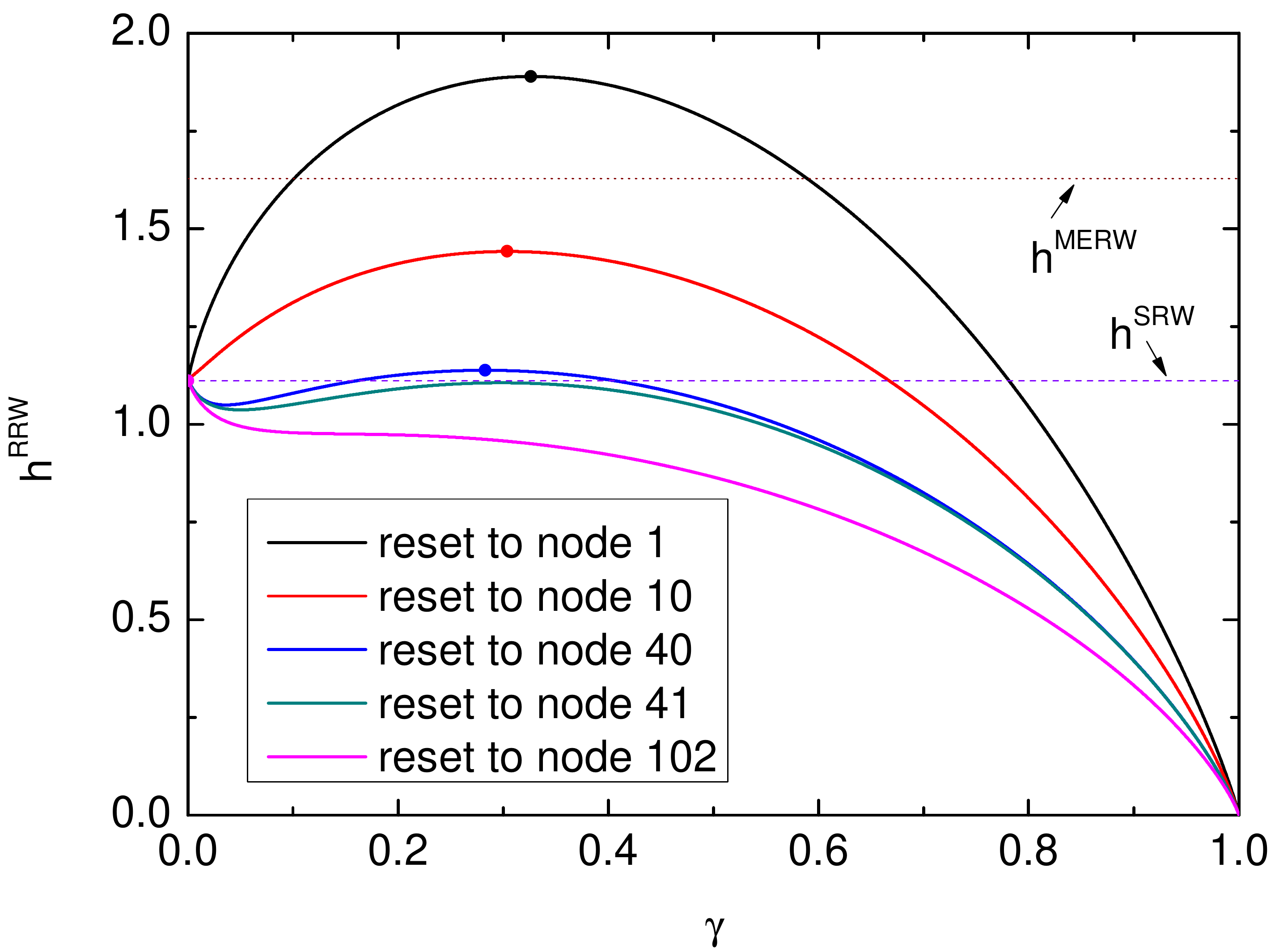}}
	\caption{Entropy rate $h^{RRW}$ of resetting random walks as a function of the resetting probability $\gamma$ on a BA network shown in Fig.\ref{fig3}. The symbols indicate the maximum entropy rate $h_{\max}^{RRW}$. Different lines correspond to the cases where different nodes is chosen as the resetting node, respectively. The upper and lower horizontal lines indicate the values of entropy rate for the MERW and SRW, $h^{MERW}$ and $h^{SRW}$, respectively.    \label{fig4}}
\end{figure}

As the second example, we consider a Cayley tree $C_{b,n}$, where $b$ is the coordination number except for the outermost nodes and $n$ is the number of shells. The network is generated as follows. Initially ($n = 0$), $C_{b,0}$ consists of only a central node. To form $C_{b, 1}$, $b$ nodes are created and are attached to the central node. For any $n > 1$, $C_{b, n}$ is obtained from $C_{b, n-1}$ by performing the following operation. For each boundary node of $C_{b, n-1}$, $b-1$ nodes are generated and attached to the boundary node. The size of Cayley tree is $N=1+b(2^n-1)$. In Fig.\ref{fig2}, we show the entropy rate $h^{RRW}$ as a function of $\gamma$ on a Cayley tree $C_{3,5}$ (see the inset of Fig.\ref{fig2}). The different lines represent the cases when the node from different shells is chosen as the resetting node. The upper and lower horizontal lines indicate the values of entropy rate of MERW and SRW, respectively. From Fig.\ref{fig2}, one can see that for all cases $h^{RRW}$ reaches a maximum value at a nonzero value of $\gamma$. When the resetting node is located at the inner layer, $h^{RRW}$ shows a larger value for any $\gamma$, except for $\gamma=0$ and $\gamma=1$ where $h^{RRW}$ coincides with $h^{SRW}$ and zero, respectively. Compared with the entropy rate of MERW, $h^{RRW}$ can be larger than $h^{MERW}$ in a wide range of $\gamma$, especially for the case when the inner node is selected as the resetting node.  

\begin{figure}
	\centerline{\includegraphics*[width=1.0\columnwidth]{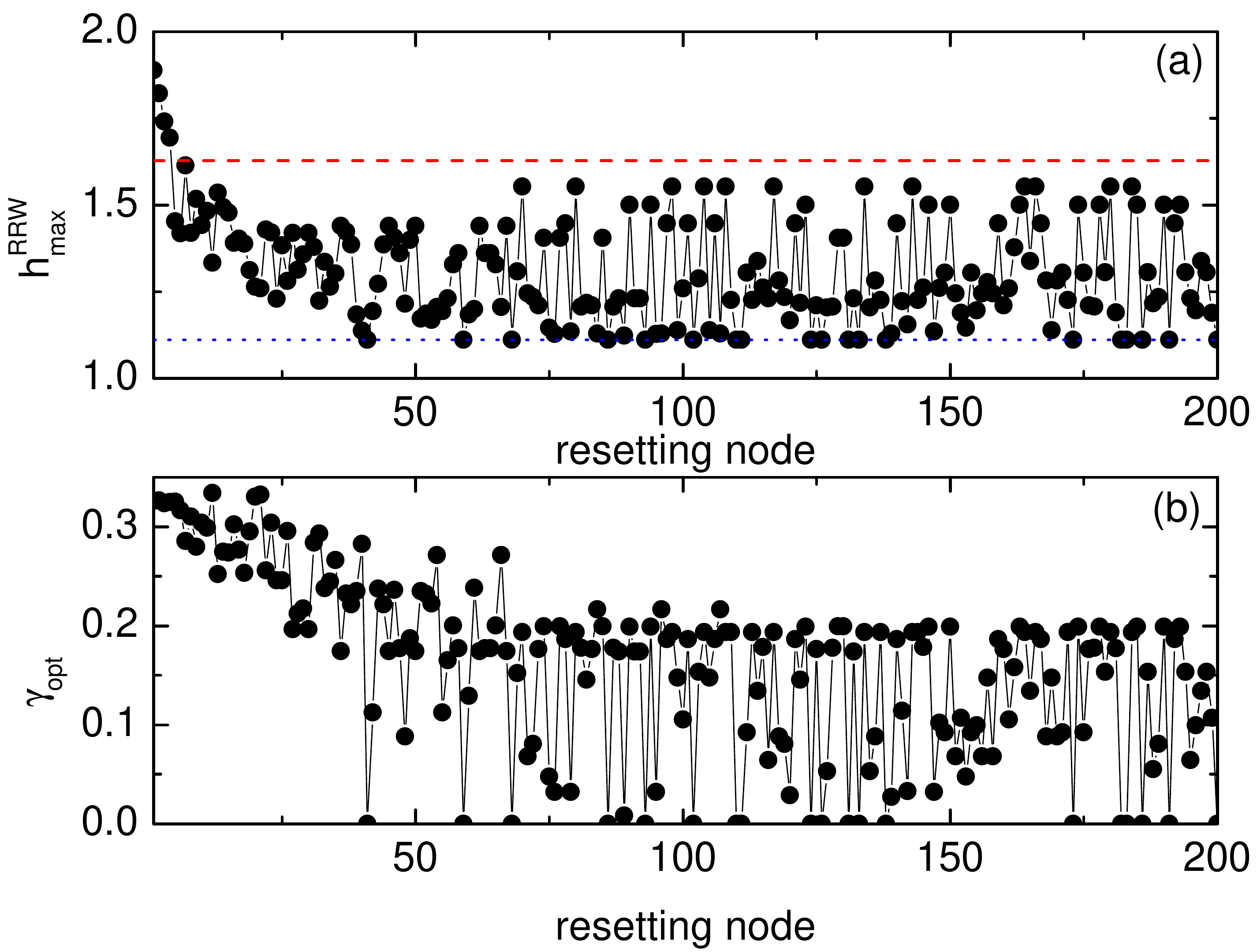}}
	\caption{The maximum entropy rate $h^{RRW}_{\max}$ (a) of resetting random walks and  the optimal resetting probability $\gamma_{{\rm{opt}}}$ (b) corresponding to $h^{RRW}_{\max}$ as a function of the resetting node label shown in Fig.\ref{fig3}. The upper and lower horizontal lines in (a) indicate the values of entropy rate for the MERW and SRW, $h^{MERW}$ and $h^{SRW}$, respectively. \label{fig5}}
\end{figure}

Finally, we consider the RRW on a Barab\'asi-Albert (BA) network \cite{Science.286.509} of size $N=200$ and average degree $\langle k \rangle =2$, as shown in Fig.\ref{fig3}. Nodes have been numbered in descending order by nodes' degrees. In Fig.\ref{fig4}, we show the entropy rate $h^{RRW}$ as a function of the resetting probability $\gamma$. $h^{RRW}$ shows more abundant behaviors with $\gamma$ when the walker is reset to different nodes. For example, when the walker is reset to node 1 or node 10, $h^{RRW}$ shows a similar behavior as in Fig.\ref{fig1} and Fig.\ref{fig2}. In this case, $h^{RRW}$ exhibits a unique maximum at a nonzero value of $\gamma$. When node 40 or node 41 is set to be the resetting node, $h^{RRW}$ shows a more complex change with $\gamma$. If node 102 is chosen as the resetting node, $h^{RRW}$ decreases monotonically as $\gamma$ increases. For each resetting node, we fix the maximum value of $h^{RRW}$, $h^{RRW}_{\max}$, and the corresponding resetting probability, $\gamma_{\rm{ opt}}$, as shown in Fig.\ref{fig5}. According to the value of $h^{RRW}_{\max}$, we can classified all nodes into three types. The first type includes four nodes with the largest degrees (see diamonds in Fig.\ref{fig3}). When one of four nodes is chosen as the resetting node, $h^{RRW}_{\max}>h^{MERW}$ and $\gamma_{\rm{ opt}}$ lies between 0.31 and 0.33. While the walker is reset to any of other nodes, $h^{RRW}$ is always less than $h^{MERW}$ for any value of $\gamma$. Among them, there are 15 nodes (see triangles in Fig.\ref{fig3}) for which $h^{RRW}$ is even less than $h^{SRW}$ for any nonzero value of $\gamma$. We can see that the these 15 nodes are located at the periphery of the network. While for the third type of nodes (see circles in Fig.\ref{fig3}) is chosen as the resetting node, i.e., most of nodes in the network, we have $h^{SRW}<h^{RRW}_{\max}<h^{MERW}$.


\section{Conclusions}
In conclusion, we have studied the entropy rate of random walks on complex networks subject to stochastic resetting to a given node with a constant probability $\gamma$. We have computed the entropy rate $h^{RRW}$ of the resetting random walks on three different types of networks. For the $k$-regular random networks, we have shown that $h^{RRW}$ is a nonmonotonic function of $\gamma$, and proved that $h^{RRW}$ admits a maximum $h^{RRW}_{\max}=\ln (k+1)$ at $\gamma =\frac{1}{k+1}$. It is worth noting that $h^{RRW}$ is larger than that of SRW or MERW, that is $h^{SRW}=h^{MERW}=\ln k$, for $0<\gamma<\gamma_c$, where $\gamma_c$ is determined by a transcendental equation $k\gamma_c  = {\left( {1 - \gamma_c } \right)^{\frac{{\gamma_c  - 1}}{\gamma_c }}}$. Subsequently, we consider $h^{RRW}$ on a Cayley tree $C_{3,5}$. No matter which shell of nodes is chosen as the resetting node, $h^{RRW}$ exhibits also nonmonotonic dependence on $\gamma$. A maximum entropy rate occurs at a nonzero value of $\gamma$. Such a maximum entropy rate is larger than that of SRW and MERW. When the walker is reset to the inner shell, the entropy rate becomes larger. Finally, we consider a degree heterogeneous network, i.e., a BA network of size $N=200$ and average degree $\langle k \rangle =2$. We find that the dependence of $h^{RRW}$ on $\gamma$ is more complex, and it highly depends on the resetting node. When the resetting node is one of these nodes with largest degrees, $h^{RRW}$ has a unique maximum as well, and the maximum $h^{RRW}$ is larger than $h^{MERW}$. While the periphery nodes are set to be the resetting node, $h^{RRW}$ decreases monotonically with $\gamma$ such that $h^{RRW}<h^{SRW}<h^{MERW}$. When the walker is reset to one of remaining nodes, which dominate a majority of proportion, the maximum entropy rate is between $h^{SRW}$ and $h^{MERW}$. 
   
The concept of entropy rate and its maximization can find its applications to information dissemination in social networks, data packet delivery in compute networks, or to the design of efficient vaccination campaigns.  Our results indicate that it is possible to maximize the entropy rate on a given topology by a rather simple resetting operation. Therefore, our findings provide an additional story about nontivial effects of stochastic resetting in the very active field.

\begin{acknowledgments}
	This work was supported by the National Natural Science Foundation of China (11875069, 61973001).
\end{acknowledgments}

\begin{thebibliography}{81}
	\expandafter\ifx\csname natexlab\endcsname\relax\def\natexlab#1{#1}\fi
	\expandafter\ifx\csname bibnamefont\endcsname\relax
	\def\bibnamefont#1{#1}\fi
	\expandafter\ifx\csname bibfnamefont\endcsname\relax
	\def\bibfnamefont#1{#1}\fi
	\expandafter\ifx\csname citenamefont\endcsname\relax
	\def\citenamefont#1{#1}\fi
	\expandafter\ifx\csname url\endcsname\relax
	\def\url#1{\texttt{#1}}\fi
	\expandafter\ifx\csname urlprefix\endcsname\relax\def\urlprefix{URL }\fi
	\providecommand{\bibinfo}[2]{#2}
	\providecommand{\eprint}[2][]{\url{#2}}
	
	\bibitem[{\citenamefont{Evans et~al.}(2020)\citenamefont{Evans, Majumdar, and
			Schehr}}]{evans2020stochastic}
	\bibinfo{author}{\bibfnamefont{M.~R.} \bibnamefont{Evans}},
	\bibinfo{author}{\bibfnamefont{S.~N.} \bibnamefont{Majumdar}},
	\bibnamefont{and} \bibinfo{author}{\bibfnamefont{G.}~\bibnamefont{Schehr}},
	\bibinfo{journal}{J. Phys. A: Math. Theor.} \textbf{\bibinfo{volume}{53}},
	\bibinfo{pages}{193001} (\bibinfo{year}{2020}).
	
	\bibitem[{\citenamefont{B\'enichou et~al.}(2011)\citenamefont{B\'enichou,
			Loverdo, Moreau, and Voituriez}}]{RevModPhys.83.81}
	\bibinfo{author}{\bibfnamefont{O.}~\bibnamefont{B\'enichou}},
	\bibinfo{author}{\bibfnamefont{C.}~\bibnamefont{Loverdo}},
	\bibinfo{author}{\bibfnamefont{M.}~\bibnamefont{Moreau}}, \bibnamefont{and}
	\bibinfo{author}{\bibfnamefont{R.}~\bibnamefont{Voituriez}},
	\bibinfo{journal}{Rev. Mod. Phys.} \textbf{\bibinfo{volume}{83}},
	\bibinfo{pages}{81} (\bibinfo{year}{2011}).
	
	\bibitem[{\citenamefont{Boyer and Solis-Salas}(2014)}]{PhysRevLett.112.240601}
	\bibinfo{author}{\bibfnamefont{D.}~\bibnamefont{Boyer}} \bibnamefont{and}
	\bibinfo{author}{\bibfnamefont{C.}~\bibnamefont{Solis-Salas}},
	\bibinfo{journal}{Phys. Rev. Lett.} \textbf{\bibinfo{volume}{112}},
	\bibinfo{pages}{240601} (\bibinfo{year}{2014}).
	
	\bibitem[{\citenamefont{Evans et~al.}(2022)\citenamefont{Evans, Majumdar, and
			Schehr}}]{JPA2022.55.274005}
	\bibinfo{author}{\bibfnamefont{M.~R.} \bibnamefont{Evans}},
	\bibinfo{author}{\bibfnamefont{S.~N.} \bibnamefont{Majumdar}},
	\bibnamefont{and} \bibinfo{author}{\bibfnamefont{G.}~\bibnamefont{Schehr}},
	\bibinfo{journal}{J. Phys. A: Math. Theor.} \textbf{\bibinfo{volume}{55}},
	\bibinfo{pages}{274005} (\bibinfo{year}{2022}).
	
	\bibitem[{\citenamefont{Vilk et~al.}(2022)\citenamefont{Vilk, Campos, M\'endez,
			Lourie, Nathan, and Assaf}}]{PhysRevLett.128.148301}
	\bibinfo{author}{\bibfnamefont{O.}~\bibnamefont{Vilk}},
	\bibinfo{author}{\bibfnamefont{D.}~\bibnamefont{Campos}},
	\bibinfo{author}{\bibfnamefont{V.}~\bibnamefont{M\'endez}},
	\bibinfo{author}{\bibfnamefont{E.}~\bibnamefont{Lourie}},
	\bibinfo{author}{\bibfnamefont{R.}~\bibnamefont{Nathan}}, \bibnamefont{and}
	\bibinfo{author}{\bibfnamefont{M.}~\bibnamefont{Assaf}},
	\bibinfo{journal}{Phys. Rev. Lett.} \textbf{\bibinfo{volume}{128}},
	\bibinfo{pages}{148301} (\bibinfo{year}{2022}).
	
	\bibitem[{\citenamefont{Luby et~al.}(1993)\citenamefont{Luby, Sinclair, and
			Zuckerman}}]{Luby1993}
	\bibinfo{author}{\bibfnamefont{M.}~\bibnamefont{Luby}},
	\bibinfo{author}{\bibfnamefont{A.}~\bibnamefont{Sinclair}}, \bibnamefont{and}
	\bibinfo{author}{\bibfnamefont{D.}~\bibnamefont{Zuckerman}},
	\bibinfo{journal}{Inf. Process. Lett.} \textbf{\bibinfo{volume}{47}},
	\bibinfo{pages}{173} (\bibinfo{year}{1993}).
	
	\bibitem[{\citenamefont{Montanari and Zecchina}(2002)}]{PhysRevLett.88.178701}
	\bibinfo{author}{\bibfnamefont{A.}~\bibnamefont{Montanari}} \bibnamefont{and}
	\bibinfo{author}{\bibfnamefont{R.}~\bibnamefont{Zecchina}},
	\bibinfo{journal}{Phys. Rev. Lett.} \textbf{\bibinfo{volume}{88}},
	\bibinfo{pages}{178701} (\bibinfo{year}{2002}).
	
	\bibitem[{\citenamefont{P et~al.}(2010)\citenamefont{P, Allen, Majumdar, and
			Evans}}]{Biophys.J.2010.98.1099}
	\bibinfo{author}{\bibfnamefont{P.~V.} \bibnamefont{P}},
	\bibinfo{author}{\bibfnamefont{R.~J.} \bibnamefont{Allen}},
	\bibinfo{author}{\bibfnamefont{S.~N.} \bibnamefont{Majumdar}},
	\bibnamefont{and} \bibinfo{author}{\bibfnamefont{M.~R.} \bibnamefont{Evans}},
	\bibinfo{journal}{Biophys. J.} \textbf{\bibinfo{volume}{98}},
	\bibinfo{pages}{1099} (\bibinfo{year}{2010}).
	
	\bibitem[{\citenamefont{Evans and
			Majumdar}(2011{\natexlab{a}})}]{evans2011diffusion}
	\bibinfo{author}{\bibfnamefont{M.~R.} \bibnamefont{Evans}} \bibnamefont{and}
	\bibinfo{author}{\bibfnamefont{S.~N.} \bibnamefont{Majumdar}},
	\bibinfo{journal}{Phys. Rev. Lett.} \textbf{\bibinfo{volume}{106}},
	\bibinfo{pages}{160601} (\bibinfo{year}{2011}{\natexlab{a}}).
	
	\bibitem[{\citenamefont{Gupta and Jayannavar}(2022)}]{Gupta2022Review}
	\bibinfo{author}{\bibfnamefont{S.}~\bibnamefont{Gupta}} \bibnamefont{and}
	\bibinfo{author}{\bibfnamefont{A.~M.} \bibnamefont{Jayannavar}},
	\bibinfo{journal}{Front. Phys.} \textbf{\bibinfo{volume}{10}},
	\bibinfo{pages}{789097} (\bibinfo{year}{2022}).
	
	\bibitem[{\citenamefont{Reuveni}(2016)}]{PhysRevLett.116.170601}
	\bibinfo{author}{\bibfnamefont{S.}~\bibnamefont{Reuveni}},
	\bibinfo{journal}{Phys. Rev. Lett.} \textbf{\bibinfo{volume}{116}},
	\bibinfo{pages}{170601} (\bibinfo{year}{2016}).
	
	\bibitem[{\citenamefont{Pal and Reuveni}(2017)}]{pal2017first}
	\bibinfo{author}{\bibfnamefont{A.}~\bibnamefont{Pal}} \bibnamefont{and}
	\bibinfo{author}{\bibfnamefont{S.}~\bibnamefont{Reuveni}},
	\bibinfo{journal}{Phys. Rev. Lett.} \textbf{\bibinfo{volume}{118}},
	\bibinfo{pages}{030603} (\bibinfo{year}{2017}).
	
	\bibitem[{\citenamefont{Pal et~al.}(2022)\citenamefont{Pal, Kostinski, and
			Reuveni}}]{JPA2022.55.021001}
	\bibinfo{author}{\bibfnamefont{A.}~\bibnamefont{Pal}},
	\bibinfo{author}{\bibfnamefont{S.}~\bibnamefont{Kostinski}},
	\bibnamefont{and} \bibinfo{author}{\bibfnamefont{S.}~\bibnamefont{Reuveni}},
	\bibinfo{journal}{J. Phys. A: Math. Theor.} \textbf{\bibinfo{volume}{55}},
	\bibinfo{pages}{021001} (\bibinfo{year}{2022}).
	
	\bibitem[{\citenamefont{Chechkin and Sokolov}(2018)}]{chechkin2018random}
	\bibinfo{author}{\bibfnamefont{A.}~\bibnamefont{Chechkin}} \bibnamefont{and}
	\bibinfo{author}{\bibfnamefont{I.}~\bibnamefont{Sokolov}},
	\bibinfo{journal}{Phys. Rev. Lett.} \textbf{\bibinfo{volume}{121}},
	\bibinfo{pages}{050601} (\bibinfo{year}{2018}).
	
	\bibitem[{\citenamefont{Evans and
			Majumdar}(2011{\natexlab{b}})}]{evans2011diffusion2}
	\bibinfo{author}{\bibfnamefont{M.~R.} \bibnamefont{Evans}} \bibnamefont{and}
	\bibinfo{author}{\bibfnamefont{S.~N.} \bibnamefont{Majumdar}},
	\bibinfo{journal}{J. Phys. A: Math. Theor.} \textbf{\bibinfo{volume}{44}},
	\bibinfo{pages}{435001} (\bibinfo{year}{2011}{\natexlab{b}}).
	
	\bibitem[{\citenamefont{Kusmierz et~al.}(2014)\citenamefont{Kusmierz, Majumdar,
			Sabhapandit, and Schehr}}]{PhysRevLett.113.220602}
	\bibinfo{author}{\bibfnamefont{L.}~\bibnamefont{Kusmierz}},
	\bibinfo{author}{\bibfnamefont{S.~N.} \bibnamefont{Majumdar}},
	\bibinfo{author}{\bibfnamefont{S.}~\bibnamefont{Sabhapandit}},
	\bibnamefont{and} \bibinfo{author}{\bibfnamefont{G.}~\bibnamefont{Schehr}},
	\bibinfo{journal}{Phys. Rev. Lett.} \textbf{\bibinfo{volume}{113}},
	\bibinfo{pages}{220602} (\bibinfo{year}{2014}).
	
	\bibitem[{\citenamefont{Ahmad et~al.}(2019)\citenamefont{Ahmad, Nayak, Bansal,
			Nandi, and Das}}]{ahmad2019first}
	\bibinfo{author}{\bibfnamefont{S.}~\bibnamefont{Ahmad}},
	\bibinfo{author}{\bibfnamefont{I.}~\bibnamefont{Nayak}},
	\bibinfo{author}{\bibfnamefont{A.}~\bibnamefont{Bansal}},
	\bibinfo{author}{\bibfnamefont{A.}~\bibnamefont{Nandi}}, \bibnamefont{and}
	\bibinfo{author}{\bibfnamefont{D.}~\bibnamefont{Das}},
	\bibinfo{journal}{Phys. Rev. E} \textbf{\bibinfo{volume}{99}},
	\bibinfo{pages}{022130} (\bibinfo{year}{2019}).
	
	\bibitem[{\citenamefont{Eule and Metzger}(2016)}]{NJP2016.18.033006}
	\bibinfo{author}{\bibfnamefont{S.}~\bibnamefont{Eule}} \bibnamefont{and}
	\bibinfo{author}{\bibfnamefont{J.~J.} \bibnamefont{Metzger}},
	\bibinfo{journal}{New J. Phys.} \textbf{\bibinfo{volume}{18}},
	\bibinfo{pages}{033006} (\bibinfo{year}{2016}).
	
	\bibitem[{\citenamefont{Pal et~al.}(2016)\citenamefont{Pal, Kundu, and
			Evans}}]{pal2016diffusion}
	\bibinfo{author}{\bibfnamefont{A.}~\bibnamefont{Pal}},
	\bibinfo{author}{\bibfnamefont{A.}~\bibnamefont{Kundu}}, \bibnamefont{and}
	\bibinfo{author}{\bibfnamefont{M.~R.} \bibnamefont{Evans}},
	\bibinfo{journal}{J. Phys. A: Math. Theor.} \textbf{\bibinfo{volume}{49}},
	\bibinfo{pages}{225001} (\bibinfo{year}{2016}).
	
	\bibitem[{\citenamefont{Nagar and Gupta}(2016)}]{PhysRevE.93.060102}
	\bibinfo{author}{\bibfnamefont{A.}~\bibnamefont{Nagar}} \bibnamefont{and}
	\bibinfo{author}{\bibfnamefont{S.}~\bibnamefont{Gupta}},
	\bibinfo{journal}{Phys. Rev. E} \textbf{\bibinfo{volume}{93}},
	\bibinfo{pages}{060102} (\bibinfo{year}{2016}).
	
	\bibitem[{\citenamefont{Shkilev}(2017)}]{PhysRevE.96.012126}
	\bibinfo{author}{\bibfnamefont{V.~P.} \bibnamefont{Shkilev}},
	\bibinfo{journal}{Phys. Rev. E} \textbf{\bibinfo{volume}{96}},
	\bibinfo{pages}{012126} (\bibinfo{year}{2017}).
	
	\bibitem[{\citenamefont{Evans and Majumdar}(2014)}]{evans2014diffusion}
	\bibinfo{author}{\bibfnamefont{M.~R.} \bibnamefont{Evans}} \bibnamefont{and}
	\bibinfo{author}{\bibfnamefont{S.~N.} \bibnamefont{Majumdar}},
	\bibinfo{journal}{J. Phys. A: Math. Theor.} \textbf{\bibinfo{volume}{47}},
	\bibinfo{pages}{285001} (\bibinfo{year}{2014}).
	
	\bibitem[{\citenamefont{Evans and Majumdar}(2018)}]{evans2018run}
	\bibinfo{author}{\bibfnamefont{M.~R.} \bibnamefont{Evans}} \bibnamefont{and}
	\bibinfo{author}{\bibfnamefont{S.~N.} \bibnamefont{Majumdar}},
	\bibinfo{journal}{J. Phys. A: Math. Theor.} \textbf{\bibinfo{volume}{51}},
	\bibinfo{pages}{475003} (\bibinfo{year}{2018}).
	
	\bibitem[{\citenamefont{Kumar et~al.}(2020)\citenamefont{Kumar, Sadekar, and
			Basu}}]{kumar2020active}
	\bibinfo{author}{\bibfnamefont{V.}~\bibnamefont{Kumar}},
	\bibinfo{author}{\bibfnamefont{O.}~\bibnamefont{Sadekar}}, \bibnamefont{and}
	\bibinfo{author}{\bibfnamefont{U.}~\bibnamefont{Basu}},
	\bibinfo{journal}{Phys. Rev. E} \textbf{\bibinfo{volume}{102}},
	\bibinfo{pages}{052129} (\bibinfo{year}{2020}).
	
	\bibitem[{\citenamefont{Majumdar et~al.}(2021)\citenamefont{Majumdar, Mori,
			Schawe, and Schehr}}]{PhysRevE.103.022135}
	\bibinfo{author}{\bibfnamefont{S.~N.} \bibnamefont{Majumdar}},
	\bibinfo{author}{\bibfnamefont{F.}~\bibnamefont{Mori}},
	\bibinfo{author}{\bibfnamefont{H.}~\bibnamefont{Schawe}}, \bibnamefont{and}
	\bibinfo{author}{\bibfnamefont{G.}~\bibnamefont{Schehr}},
	\bibinfo{journal}{Phys. Rev. E} \textbf{\bibinfo{volume}{103}},
	\bibinfo{pages}{022135} (\bibinfo{year}{2021}).
	
	\bibitem[{\citenamefont{De~Bruyne et~al.}(2020)\citenamefont{De~Bruyne,
			Randon-Furling, and Redner}}]{de2020optimization}
	\bibinfo{author}{\bibfnamefont{B.}~\bibnamefont{De~Bruyne}},
	\bibinfo{author}{\bibfnamefont{J.}~\bibnamefont{Randon-Furling}},
	\bibnamefont{and} \bibinfo{author}{\bibfnamefont{S.}~\bibnamefont{Redner}},
	\bibinfo{journal}{Phys. Rev. Lett.} \textbf{\bibinfo{volume}{125}},
	\bibinfo{pages}{050602} (\bibinfo{year}{2020}).
	
	\bibitem[{\citenamefont{Lauber~Bonomo and Pal}(2021)}]{lauber2021first}
	\bibinfo{author}{\bibfnamefont{O.}~\bibnamefont{Lauber~Bonomo}}
	\bibnamefont{and} \bibinfo{author}{\bibfnamefont{A.}~\bibnamefont{Pal}},
	\bibinfo{journal}{arXiv: 2102.00895}  (\bibinfo{year}{2021}).
	
	\bibitem[{\citenamefont{Schumm and Bressloff}(2021)}]{JPA2021.54.404004}
	\bibinfo{author}{\bibfnamefont{R.~C.} \bibnamefont{Schumm}} \bibnamefont{and}
	\bibinfo{author}{\bibfnamefont{P.~C.} \bibnamefont{Bressloff}},
	\bibinfo{journal}{J. Phys. A: Math. Theor.} \textbf{\bibinfo{volume}{54}},
	\bibinfo{pages}{404004} (\bibinfo{year}{2021}).
	
	\bibitem[{\citenamefont{Bressloff}(2022)}]{JPA2022.55.275002}
	\bibinfo{author}{\bibfnamefont{P.~C.} \bibnamefont{Bressloff}},
	\bibinfo{journal}{J. Phys. A: Math. Theor.} \textbf{\bibinfo{volume}{55}},
	\bibinfo{pages}{275002} (\bibinfo{year}{2022}).
	
	\bibitem[{\citenamefont{Klinger et~al.}(2021)\citenamefont{Klinger, Voituriez,
			and B\'enichou}}]{PhysRevE.103.032107}
	\bibinfo{author}{\bibfnamefont{J.}~\bibnamefont{Klinger}},
	\bibinfo{author}{\bibfnamefont{R.}~\bibnamefont{Voituriez}},
	\bibnamefont{and}
	\bibinfo{author}{\bibfnamefont{O.}~\bibnamefont{B\'enichou}},
	\bibinfo{journal}{Phys. Rev. E} \textbf{\bibinfo{volume}{103}},
	\bibinfo{pages}{032107} (\bibinfo{year}{2021}).
	
	\bibitem[{\citenamefont{Biroli et~al.}(2022)\citenamefont{Biroli, Mori, and
			Majumdar}}]{arXiv:2202.04906}
	\bibinfo{author}{\bibfnamefont{M.}~\bibnamefont{Biroli}},
	\bibinfo{author}{\bibfnamefont{F.}~\bibnamefont{Mori}}, \bibnamefont{and}
	\bibinfo{author}{\bibfnamefont{S.~N.} \bibnamefont{Majumdar}},
	\bibinfo{journal}{J. Phys. A: Math. Theor.} \textbf{\bibinfo{volume}{55}},
	\bibinfo{pages}{244001} (\bibinfo{year}{2022}).
	
	\bibitem[{\citenamefont{Singh and Pal}(2022)}]{JPA2022.55.234001}
	\bibinfo{author}{\bibfnamefont{P.}~\bibnamefont{Singh}} \bibnamefont{and}
	\bibinfo{author}{\bibfnamefont{A.}~\bibnamefont{Pal}}, \bibinfo{journal}{J.
		Phys. A: Math. Theor.} \textbf{\bibinfo{volume}{55}}, \bibinfo{pages}{234001}
	(\bibinfo{year}{2022}).
	
	\bibitem[{\citenamefont{Chen and Huang}(2022)}]{PhysRevE.105.034109}
	\bibinfo{author}{\bibfnamefont{H.}~\bibnamefont{Chen}} \bibnamefont{and}
	\bibinfo{author}{\bibfnamefont{F.}~\bibnamefont{Huang}},
	\bibinfo{journal}{Phys. Rev. E} \textbf{\bibinfo{volume}{105}},
	\bibinfo{pages}{034109} (\bibinfo{year}{2022}).
	
	\bibitem[{\citenamefont{Huang and Chen}(2021)}]{huang2021random}
	\bibinfo{author}{\bibfnamefont{F.}~\bibnamefont{Huang}} \bibnamefont{and}
	\bibinfo{author}{\bibfnamefont{H.}~\bibnamefont{Chen}},
	\bibinfo{journal}{Phys. Rev. E} \textbf{\bibinfo{volume}{103}},
	\bibinfo{pages}{062132} (\bibinfo{year}{2021}).
	
	\bibitem[{\citenamefont{De~Bruyne et~al.}(2022)\citenamefont{De~Bruyne,
			Majumdar, and Schehr}}]{PhysRevLett.128.200603}
	\bibinfo{author}{\bibfnamefont{B.}~\bibnamefont{De~Bruyne}},
	\bibinfo{author}{\bibfnamefont{S.~N.} \bibnamefont{Majumdar}},
	\bibnamefont{and} \bibinfo{author}{\bibfnamefont{G.}~\bibnamefont{Schehr}},
	\bibinfo{journal}{Phys. Rev. Lett.} \textbf{\bibinfo{volume}{128}},
	\bibinfo{pages}{200603} (\bibinfo{year}{2022}).
	
	\bibitem[{\citenamefont{Gupta et~al.}(2014)\citenamefont{Gupta, Majumdar, and
			Schehr}}]{gupta2014fluctuating}
	\bibinfo{author}{\bibfnamefont{S.}~\bibnamefont{Gupta}},
	\bibinfo{author}{\bibfnamefont{S.~N.} \bibnamefont{Majumdar}},
	\bibnamefont{and} \bibinfo{author}{\bibfnamefont{G.}~\bibnamefont{Schehr}},
	\bibinfo{journal}{Phys. Rev. Lett.} \textbf{\bibinfo{volume}{112}},
	\bibinfo{pages}{220601} (\bibinfo{year}{2014}).
	
	\bibitem[{\citenamefont{Fuchs et~al.}(2016)\citenamefont{Fuchs, Goldt, and
			Seifert}}]{fuchs2016stochastic}
	\bibinfo{author}{\bibfnamefont{J.}~\bibnamefont{Fuchs}},
	\bibinfo{author}{\bibfnamefont{S.}~\bibnamefont{Goldt}}, \bibnamefont{and}
	\bibinfo{author}{\bibfnamefont{U.}~\bibnamefont{Seifert}},
	\bibinfo{journal}{EPL (Europhys. Lett.)} \textbf{\bibinfo{volume}{113}},
	\bibinfo{pages}{60009} (\bibinfo{year}{2016}).
	
	\bibitem[{\citenamefont{Pal and Rahav}(2017)}]{pal2017integral}
	\bibinfo{author}{\bibfnamefont{A.}~\bibnamefont{Pal}} \bibnamefont{and}
	\bibinfo{author}{\bibfnamefont{S.}~\bibnamefont{Rahav}},
	\bibinfo{journal}{Phys. Rev. E} \textbf{\bibinfo{volume}{96}},
	\bibinfo{pages}{062135} (\bibinfo{year}{2017}).
	
	\bibitem[{\citenamefont{Gupta et~al.}(2020)\citenamefont{Gupta, Plata, and
			Pal}}]{gupta2020work}
	\bibinfo{author}{\bibfnamefont{D.}~\bibnamefont{Gupta}},
	\bibinfo{author}{\bibfnamefont{C.~A.} \bibnamefont{Plata}}, \bibnamefont{and}
	\bibinfo{author}{\bibfnamefont{A.}~\bibnamefont{Pal}},
	\bibinfo{journal}{Phys. Rev. Lett.} \textbf{\bibinfo{volume}{124}},
	\bibinfo{pages}{110608} (\bibinfo{year}{2020}).
	
	\bibitem[{\citenamefont{Reuveni et~al.}(2014)\citenamefont{Reuveni, Urbakh, and
			Klafter}}]{reuveni2014role}
	\bibinfo{author}{\bibfnamefont{S.}~\bibnamefont{Reuveni}},
	\bibinfo{author}{\bibfnamefont{M.}~\bibnamefont{Urbakh}}, \bibnamefont{and}
	\bibinfo{author}{\bibfnamefont{J.}~\bibnamefont{Klafter}},
	\bibinfo{journal}{Proc. Natl. Acad. Sci. USA} \textbf{\bibinfo{volume}{111}},
	\bibinfo{pages}{4391} (\bibinfo{year}{2014}).
	
	\bibitem[{\citenamefont{Rotbart et~al.}(2015)\citenamefont{Rotbart, Reuveni,
			and Urbakh}}]{rotbart2015michaelis}
	\bibinfo{author}{\bibfnamefont{T.}~\bibnamefont{Rotbart}},
	\bibinfo{author}{\bibfnamefont{S.}~\bibnamefont{Reuveni}}, \bibnamefont{and}
	\bibinfo{author}{\bibfnamefont{M.}~\bibnamefont{Urbakh}},
	\bibinfo{journal}{Phys. Rev. E} \textbf{\bibinfo{volume}{92}},
	\bibinfo{pages}{060101} (\bibinfo{year}{2015}).
	
	\bibitem[{\citenamefont{Meylahn et~al.}(2015)\citenamefont{Meylahn,
			Sabhapandit, and Touchette}}]{meylahn2015large}
	\bibinfo{author}{\bibfnamefont{J.~M.} \bibnamefont{Meylahn}},
	\bibinfo{author}{\bibfnamefont{S.}~\bibnamefont{Sabhapandit}},
	\bibnamefont{and}
	\bibinfo{author}{\bibfnamefont{H.}~\bibnamefont{Touchette}},
	\bibinfo{journal}{Phys. Rev. E} \textbf{\bibinfo{volume}{92}},
	\bibinfo{pages}{062148} (\bibinfo{year}{2015}).
	
	\bibitem[{\citenamefont{Singh and Pal}(2021)}]{PhysRevE.103.052119}
	\bibinfo{author}{\bibfnamefont{P.}~\bibnamefont{Singh}} \bibnamefont{and}
	\bibinfo{author}{\bibfnamefont{A.}~\bibnamefont{Pal}},
	\bibinfo{journal}{Phys. Rev. E} \textbf{\bibinfo{volume}{103}},
	\bibinfo{pages}{052119} (\bibinfo{year}{2021}).
	
	\bibitem[{\citenamefont{Godr\'eche and Luck}(2022)}]{JStatMech2022.063202}
	\bibinfo{author}{\bibfnamefont{C.}~\bibnamefont{Godr\'eche}} \bibnamefont{and}
	\bibinfo{author}{\bibfnamefont{J.-M.} \bibnamefont{Luck}},
	\bibinfo{journal}{J. Stat. Mech.} \textbf{\bibinfo{volume}{2022}},
	\bibinfo{pages}{063202} (\bibinfo{year}{2022}).
	
	\bibitem[{\citenamefont{Majumdar et~al.}(2022)\citenamefont{Majumdar, Mounaix,
			Sabhapandit, and Schehr}}]{JPA2022.55.034002}
	\bibinfo{author}{\bibfnamefont{S.~N.} \bibnamefont{Majumdar}},
	\bibinfo{author}{\bibfnamefont{P.}~\bibnamefont{Mounaix}},
	\bibinfo{author}{\bibfnamefont{S.}~\bibnamefont{Sabhapandit}},
	\bibnamefont{and} \bibinfo{author}{\bibfnamefont{G.}~\bibnamefont{Schehr}},
	\bibinfo{journal}{J. Phys. A: Math. Theor.} \textbf{\bibinfo{volume}{55}},
	\bibinfo{pages}{034002} (\bibinfo{year}{2022}).
	
	\bibitem[{\citenamefont{Bruynea and Mori}(2021)}]{arXiv:2112.11416}
	\bibinfo{author}{\bibfnamefont{B.~D.} \bibnamefont{Bruynea}} \bibnamefont{and}
	\bibinfo{author}{\bibfnamefont{F.}~\bibnamefont{Mori}},
	\bibinfo{journal}{arXiv:2112.11416}  (\bibinfo{year}{2021}).
	
	\bibitem[{\citenamefont{Tal-Friedman et~al.}(2020)\citenamefont{Tal-Friedman,
			Pal, Sekhon, Reuveni, and Roichman}}]{tal2020experimental}
	\bibinfo{author}{\bibfnamefont{O.}~\bibnamefont{Tal-Friedman}},
	\bibinfo{author}{\bibfnamefont{A.}~\bibnamefont{Pal}},
	\bibinfo{author}{\bibfnamefont{A.}~\bibnamefont{Sekhon}},
	\bibinfo{author}{\bibfnamefont{S.}~\bibnamefont{Reuveni}}, \bibnamefont{and}
	\bibinfo{author}{\bibfnamefont{Y.}~\bibnamefont{Roichman}},
	\bibinfo{journal}{J. Phys. Chem. Lett.} \textbf{\bibinfo{volume}{11}},
	\bibinfo{pages}{7350} (\bibinfo{year}{2020}).
	
	\bibitem[{\citenamefont{Besga et~al.}(2020)\citenamefont{Besga, Bovon,
			Petrosyan, Majumdar, and Ciliberto}}]{besga2020optimal}
	\bibinfo{author}{\bibfnamefont{B.}~\bibnamefont{Besga}},
	\bibinfo{author}{\bibfnamefont{A.}~\bibnamefont{Bovon}},
	\bibinfo{author}{\bibfnamefont{A.}~\bibnamefont{Petrosyan}},
	\bibinfo{author}{\bibfnamefont{S.~N.} \bibnamefont{Majumdar}},
	\bibnamefont{and}
	\bibinfo{author}{\bibfnamefont{S.}~\bibnamefont{Ciliberto}},
	\bibinfo{journal}{Phys. Rev. Research} \textbf{\bibinfo{volume}{2}},
	\bibinfo{pages}{032029} (\bibinfo{year}{2020}).
	
	\bibitem[{\citenamefont{Bianconi}(2008)}]{Bianconi_EPL2028}
	\bibinfo{author}{\bibfnamefont{G.}~\bibnamefont{Bianconi}},
	\bibinfo{journal}{EPL} \textbf{\bibinfo{volume}{81}}, \bibinfo{pages}{28005}
	(\bibinfo{year}{2008}).
	
	\bibitem[{\citenamefont{Bianconi}(2009)}]{PhysRevE.79.036114}
	\bibinfo{author}{\bibfnamefont{G.}~\bibnamefont{Bianconi}},
	\bibinfo{journal}{Phys. Rev. E} \textbf{\bibinfo{volume}{79}},
	\bibinfo{pages}{036114} (\bibinfo{year}{2009}).
	
	\bibitem[{\citenamefont{Anand and Bianconi}(2009)}]{PhysRevE.80.045102}
	\bibinfo{author}{\bibfnamefont{K.}~\bibnamefont{Anand}} \bibnamefont{and}
	\bibinfo{author}{\bibfnamefont{G.}~\bibnamefont{Bianconi}},
	\bibinfo{journal}{Phys. Rev. E} \textbf{\bibinfo{volume}{80}},
	\bibinfo{pages}{045102} (\bibinfo{year}{2009}).
	
	\bibitem[{\citenamefont{Anand and Bianconi}(2010)}]{PhysRevE.82.011116}
	\bibinfo{author}{\bibfnamefont{K.}~\bibnamefont{Anand}} \bibnamefont{and}
	\bibinfo{author}{\bibfnamefont{G.}~\bibnamefont{Bianconi}},
	\bibinfo{journal}{Phys. Rev. E} \textbf{\bibinfo{volume}{82}},
	\bibinfo{pages}{011116} (\bibinfo{year}{2010}).
	
	\bibitem[{\citenamefont{Park and Newman}(2004)}]{PhysRevE.70.066117}
	\bibinfo{author}{\bibfnamefont{J.}~\bibnamefont{Park}} \bibnamefont{and}
	\bibinfo{author}{\bibfnamefont{M.~E.~J.} \bibnamefont{Newman}},
	\bibinfo{journal}{Phys. Rev. E} \textbf{\bibinfo{volume}{70}},
	\bibinfo{pages}{066117} (\bibinfo{year}{2004}).
	
	\bibitem[{\citenamefont{Squartini and Garlaschelli}(2011)}]{NJP2011.13.083001}
	\bibinfo{author}{\bibfnamefont{T.}~\bibnamefont{Squartini}} \bibnamefont{and}
	\bibinfo{author}{\bibfnamefont{D.}~\bibnamefont{Garlaschelli}},
	\bibinfo{journal}{New J. Phys.} \textbf{\bibinfo{volume}{13}},
	\bibinfo{pages}{083001} (\bibinfo{year}{2011}).
	
	\bibitem[{\citenamefont{Horv\'at et~al.}(2015)\citenamefont{Horv\'at, Czabarka,
			and Toroczkai}}]{PhysRevLett.114.158701}
	\bibinfo{author}{\bibfnamefont{S.}~\bibnamefont{Horv\'at}},
	\bibinfo{author}{\bibfnamefont{E.}~\bibnamefont{Czabarka}}, \bibnamefont{and}
	\bibinfo{author}{\bibfnamefont{Z.}~\bibnamefont{Toroczkai}},
	\bibinfo{journal}{Phys. Rev. Lett.} \textbf{\bibinfo{volume}{114}},
	\bibinfo{pages}{158701} (\bibinfo{year}{2015}).
	
	\bibitem[{\citenamefont{Squartini et~al.}(2015)\citenamefont{Squartini, de~Mol,
			den Hollander, and Garlaschelli}}]{PhysRevLett.115.268701}
	\bibinfo{author}{\bibfnamefont{T.}~\bibnamefont{Squartini}},
	\bibinfo{author}{\bibfnamefont{J.}~\bibnamefont{de~Mol}},
	\bibinfo{author}{\bibfnamefont{F.}~\bibnamefont{den Hollander}},
	\bibnamefont{and}
	\bibinfo{author}{\bibfnamefont{D.}~\bibnamefont{Garlaschelli}},
	\bibinfo{journal}{Phys. Rev. Lett.} \textbf{\bibinfo{volume}{115}},
	\bibinfo{pages}{268701} (\bibinfo{year}{2015}).
	
	\bibitem[{\citenamefont{Bianconi}(2013)}]{PhysRevE.87.062806}
	\bibinfo{author}{\bibfnamefont{G.}~\bibnamefont{Bianconi}},
	\bibinfo{journal}{Phys. Rev. E} \textbf{\bibinfo{volume}{87}},
	\bibinfo{pages}{062806} (\bibinfo{year}{2013}).
	
	\bibitem[{\citenamefont{Courtney and Bianconi}(2016)}]{PhysRevE.93.062311}
	\bibinfo{author}{\bibfnamefont{O.~T.} \bibnamefont{Courtney}} \bibnamefont{and}
	\bibinfo{author}{\bibfnamefont{G.}~\bibnamefont{Bianconi}},
	\bibinfo{journal}{Phys. Rev. E} \textbf{\bibinfo{volume}{93}},
	\bibinfo{pages}{062311} (\bibinfo{year}{2016}).
	
	\bibitem[{\citenamefont{Cimini et~al.}(2019)\citenamefont{Cimini, Squartini,
			Saracco, Garlaschelli, Gabrielli, and Caldarelli}}]{NatRevPhys2019.1.58}
	\bibinfo{author}{\bibfnamefont{G.}~\bibnamefont{Cimini}},
	\bibinfo{author}{\bibfnamefont{T.}~\bibnamefont{Squartini}},
	\bibinfo{author}{\bibfnamefont{F.}~\bibnamefont{Saracco}},
	\bibinfo{author}{\bibfnamefont{.}~\bibnamefont{Garlaschelli}},
	\bibinfo{author}{\bibfnamefont{A.}~\bibnamefont{Gabrielli}},
	\bibnamefont{and}
	\bibinfo{author}{\bibfnamefont{G.}~\bibnamefont{Caldarelli}},
	\bibinfo{journal}{Nat. Rev. Phys.} \textbf{\bibinfo{volume}{1}},
	\bibinfo{pages}{58?71} (\bibinfo{year}{2019}).
	
	\bibitem[{\citenamefont{Radicchi and
			Castellano}(2018)}]{PhysRevLett.120.198301}
	\bibinfo{author}{\bibfnamefont{F.}~\bibnamefont{Radicchi}} \bibnamefont{and}
	\bibinfo{author}{\bibfnamefont{C.}~\bibnamefont{Castellano}},
	\bibinfo{journal}{Phys. Rev. Lett.} \textbf{\bibinfo{volume}{120}},
	\bibinfo{pages}{198301} (\bibinfo{year}{2018}).
	
	\bibitem[{\citenamefont{Bianconi et~al.}(2009)\citenamefont{Bianconi, Pin, and
			Marsili}}]{Bianconi_PNAS2009}
	\bibinfo{author}{\bibfnamefont{G.}~\bibnamefont{Bianconi}},
	\bibinfo{author}{\bibfnamefont{P.}~\bibnamefont{Pin}}, \bibnamefont{and}
	\bibinfo{author}{\bibfnamefont{M.}~\bibnamefont{Marsili}},
	\bibinfo{journal}{Proc. Natl. Acad. Sci. USA} \textbf{\bibinfo{volume}{106}},
	\bibinfo{pages}{11433} (\bibinfo{year}{2009}).
	
	\bibitem[{\citenamefont{Burda et~al.}(2009)\citenamefont{Burda, Duda, Luck, and
			Waclaw}}]{PhysRevLett.102.160602}
	\bibinfo{author}{\bibfnamefont{Z.}~\bibnamefont{Burda}},
	\bibinfo{author}{\bibfnamefont{J.}~\bibnamefont{Duda}},
	\bibinfo{author}{\bibfnamefont{J.~M.} \bibnamefont{Luck}}, \bibnamefont{and}
	\bibinfo{author}{\bibfnamefont{B.}~\bibnamefont{Waclaw}},
	\bibinfo{journal}{Phys. Rev. Lett.} \textbf{\bibinfo{volume}{102}},
	\bibinfo{pages}{160602} (\bibinfo{year}{2009}).
	
	\bibitem[{\citenamefont{Parry}(1964)}]{Parry1964}
	\bibinfo{author}{\bibfnamefont{W.}~\bibnamefont{Parry}},
	\bibinfo{journal}{Trans. Amer. Math. Soc.} \textbf{\bibinfo{volume}{112}},
	\bibinfo{pages}{55} (\bibinfo{year}{1964}).
	
	\bibitem[{\citenamefont{Demetrius and Manke}(2005)}]{Demetrius2005}
	\bibinfo{author}{\bibfnamefont{L.}~\bibnamefont{Demetrius}} \bibnamefont{and}
	\bibinfo{author}{\bibfnamefont{T.}~\bibnamefont{Manke}},
	\bibinfo{journal}{Physica A} \textbf{\bibinfo{volume}{346}},
	\bibinfo{pages}{682?696} (\bibinfo{year}{2005}).
	
	\bibitem[{\citenamefont{Delvenne and Libert}(2011)}]{PhysRevE.83.046117}
	\bibinfo{author}{\bibfnamefont{J.-C.} \bibnamefont{Delvenne}} \bibnamefont{and}
	\bibinfo{author}{\bibfnamefont{A.-S.} \bibnamefont{Libert}},
	\bibinfo{journal}{Phys. Rev. E} \textbf{\bibinfo{volume}{83}},
	\bibinfo{pages}{046117} (\bibinfo{year}{2011}).
	
	\bibitem[{\citenamefont{G\'omez-Garde\~nes and
			Latora}(2008)}]{PhysRevE.78.065102}
	\bibinfo{author}{\bibfnamefont{J.}~\bibnamefont{G\'omez-Garde\~nes}}
	\bibnamefont{and} \bibinfo{author}{\bibfnamefont{V.}~\bibnamefont{Latora}},
	\bibinfo{journal}{Phys. Rev. E} \textbf{\bibinfo{volume}{78}},
	\bibinfo{pages}{065102} (\bibinfo{year}{2008}).
	
	\bibitem[{\citenamefont{Sinatra et~al.}(2011)\citenamefont{Sinatra,
			G\'omez-Garde\~nes, Lambiotte, Nicosia, and Latora}}]{PhysRevE.83.030103}
	\bibinfo{author}{\bibfnamefont{R.}~\bibnamefont{Sinatra}},
	\bibinfo{author}{\bibfnamefont{J.}~\bibnamefont{G\'omez-Garde\~nes}},
	\bibinfo{author}{\bibfnamefont{R.}~\bibnamefont{Lambiotte}},
	\bibinfo{author}{\bibfnamefont{V.}~\bibnamefont{Nicosia}}, \bibnamefont{and}
	\bibinfo{author}{\bibfnamefont{V.}~\bibnamefont{Latora}},
	\bibinfo{journal}{Phys. Rev. E} \textbf{\bibinfo{volume}{83}},
	\bibinfo{pages}{030103} (\bibinfo{year}{2011}).
	
	\bibitem[{\citenamefont{Zhao et~al.}(2011)\citenamefont{Zhao, Halu, Severini,
			and Bianconi}}]{PhysRevE.84.066113}
	\bibinfo{author}{\bibfnamefont{K.}~\bibnamefont{Zhao}},
	\bibinfo{author}{\bibfnamefont{A.}~\bibnamefont{Halu}},
	\bibinfo{author}{\bibfnamefont{S.}~\bibnamefont{Severini}}, \bibnamefont{and}
	\bibinfo{author}{\bibfnamefont{G.}~\bibnamefont{Bianconi}},
	\bibinfo{journal}{Phys. Rev. E} \textbf{\bibinfo{volume}{84}},
	\bibinfo{pages}{066113} (\bibinfo{year}{2011}).
	
	\bibitem[{\citenamefont{Press\'e et~al.}(2013)\citenamefont{Press\'e, Ghosh,
			Lee, and Dill}}]{RevModPhys.85.1115}
	\bibinfo{author}{\bibfnamefont{S.}~\bibnamefont{Press\'e}},
	\bibinfo{author}{\bibfnamefont{K.}~\bibnamefont{Ghosh}},
	\bibinfo{author}{\bibfnamefont{J.}~\bibnamefont{Lee}}, \bibnamefont{and}
	\bibinfo{author}{\bibfnamefont{K.~A.} \bibnamefont{Dill}},
	\bibinfo{journal}{Rev. Mod. Phys.} \textbf{\bibinfo{volume}{85}},
	\bibinfo{pages}{1115} (\bibinfo{year}{2013}).
	
	\bibitem[{\citenamefont{Wang et~al.}(2021)\citenamefont{Wang, Chen, and
			Huang}}]{Chaos2021_31.093135}
	\bibinfo{author}{\bibfnamefont{S.}~\bibnamefont{Wang}},
	\bibinfo{author}{\bibfnamefont{H.}~\bibnamefont{Chen}}, \bibnamefont{and}
	\bibinfo{author}{\bibfnamefont{F.}~\bibnamefont{Huang}},
	\bibinfo{journal}{Chaos} \textbf{\bibinfo{volume}{31}},
	\bibinfo{pages}{093135} (\bibinfo{year}{2021}).
	
	\bibitem[{\citenamefont{Chen et~al.}(2021)\citenamefont{Chen, Li, and
			Huang}}]{arXiv.2111.01330}
	\bibinfo{author}{\bibfnamefont{H.}~\bibnamefont{Chen}},
	\bibinfo{author}{\bibfnamefont{G.}~\bibnamefont{Li}}, \bibnamefont{and}
	\bibinfo{author}{\bibfnamefont{F.}~\bibnamefont{Huang}},
	\bibinfo{journal}{arXiv:2111.01330}  (\bibinfo{year}{2021}).
	
	\bibitem[{\citenamefont{Riascos et~al.}(2020)\citenamefont{Riascos, Boyer,
			Herringer, and Mateos}}]{PhysRevE.101.062147}
	\bibinfo{author}{\bibfnamefont{A.~P.} \bibnamefont{Riascos}},
	\bibinfo{author}{\bibfnamefont{D.}~\bibnamefont{Boyer}},
	\bibinfo{author}{\bibfnamefont{P.}~\bibnamefont{Herringer}},
	\bibnamefont{and} \bibinfo{author}{\bibfnamefont{J.~L.}
		\bibnamefont{Mateos}}, \bibinfo{journal}{Phys. Rev. E}
	\textbf{\bibinfo{volume}{101}}, \bibinfo{pages}{062147}
	(\bibinfo{year}{2020}).
	
	\bibitem[{\citenamefont{Gonz\'alez et~al.}(2021)\citenamefont{Gonz\'alez,
			Riascos, and Boyer}}]{PhysRevE.103.062126}
	\bibinfo{author}{\bibfnamefont{F.~H.} \bibnamefont{Gonz\'alez}},
	\bibinfo{author}{\bibfnamefont{A.~P.} \bibnamefont{Riascos}},
	\bibnamefont{and} \bibinfo{author}{\bibfnamefont{D.}~\bibnamefont{Boyer}},
	\bibinfo{journal}{Phys. Rev. E} \textbf{\bibinfo{volume}{103}},
	\bibinfo{pages}{062126} (\bibinfo{year}{2021}).
	
	\bibitem[{\citenamefont{Ye and Chen}(2022)}]{JSM2022.053201}
	\bibinfo{author}{\bibfnamefont{Y.}~\bibnamefont{Ye}} \bibnamefont{and}
	\bibinfo{author}{\bibfnamefont{H.}~\bibnamefont{Chen}}, \bibinfo{journal}{J.
		Stat. Mech.} \textbf{\bibinfo{volume}{2022}}, \bibinfo{pages}{053201}
	(\bibinfo{year}{2022}).
	
	\bibitem[{\citenamefont{Masuda et~al.}(2017)\citenamefont{Masuda, Porter, and
			Lambiotte}}]{masuda2017random}
	\bibinfo{author}{\bibfnamefont{N.}~\bibnamefont{Masuda}},
	\bibinfo{author}{\bibfnamefont{M.~A.} \bibnamefont{Porter}},
	\bibnamefont{and}
	\bibinfo{author}{\bibfnamefont{R.}~\bibnamefont{Lambiotte}},
	\bibinfo{journal}{Phys. Rep.} \textbf{\bibinfo{volume}{716}},
	\bibinfo{pages}{1} (\bibinfo{year}{2017}).
	
	\bibitem[{\citenamefont{Noh and Rieger}(2004)}]{PhysRevLett.92.118701}
	\bibinfo{author}{\bibfnamefont{J.~D.} \bibnamefont{Noh}} \bibnamefont{and}
	\bibinfo{author}{\bibfnamefont{H.}~\bibnamefont{Rieger}},
	\bibinfo{journal}{Phys. Rev. Lett.} \textbf{\bibinfo{volume}{92}},
	\bibinfo{pages}{118701} (\bibinfo{year}{2004}).
	
	\bibitem[{\citenamefont{Zhang et~al.}(2013)\citenamefont{Zhang, Shan, and
			Chen}}]{PhysRevE.87.012112}
	\bibinfo{author}{\bibfnamefont{Z.}~\bibnamefont{Zhang}},
	\bibinfo{author}{\bibfnamefont{T.}~\bibnamefont{Shan}}, \bibnamefont{and}
	\bibinfo{author}{\bibfnamefont{G.}~\bibnamefont{Chen}},
	\bibinfo{journal}{Phys. Rev. E} \textbf{\bibinfo{volume}{87}},
	\bibinfo{pages}{012112} (\bibinfo{year}{2013}).
	
	\bibitem[{\citenamefont{Bautista et~al.}(2019)\citenamefont{Bautista, Abry, and
			Gon\c{c}alves}}]{Bautista2019}
	\bibinfo{author}{\bibfnamefont{E.}~\bibnamefont{Bautista}},
	\bibinfo{author}{\bibfnamefont{P.}~\bibnamefont{Abry}}, \bibnamefont{and}
	\bibinfo{author}{\bibfnamefont{P.}~\bibnamefont{Gon\c{c}alves}},
	\bibinfo{journal}{Appl. Netw. Sci.} \textbf{\bibinfo{volume}{4}},
	\bibinfo{pages}{57} (\bibinfo{year}{2019}).
	
	\bibitem[{\citenamefont{Brin and Page}(1998)}]{Pagerank1998}
	\bibinfo{author}{\bibfnamefont{S.}~\bibnamefont{Brin}} \bibnamefont{and}
	\bibinfo{author}{\bibfnamefont{L.}~\bibnamefont{Page}},
	\bibinfo{journal}{Comput. Netw. ISDN Syst.} \textbf{\bibinfo{volume}{30}},
	\bibinfo{pages}{107} (\bibinfo{year}{1998}).
	
	\bibitem[{\citenamefont{Avrachenkov et~al.}(2018)\citenamefont{Avrachenkov,
			Piunovskiy, and Zhang}}]{avrachenkov2018hitting}
	\bibinfo{author}{\bibfnamefont{K.}~\bibnamefont{Avrachenkov}},
	\bibinfo{author}{\bibfnamefont{A.}~\bibnamefont{Piunovskiy}},
	\bibnamefont{and} \bibinfo{author}{\bibfnamefont{Y.}~\bibnamefont{Zhang}},
	\bibinfo{journal}{Methodology and Computing in Applied Probability}
	\textbf{\bibinfo{volume}{20}}, \bibinfo{pages}{1173} (\bibinfo{year}{2018}).
	
	\bibitem[{\citenamefont{Barab\'asi and Albert}(1999)}]{Science.286.509}
	\bibinfo{author}{\bibfnamefont{A.-L.} \bibnamefont{Barab\'asi}}
	\bibnamefont{and} \bibinfo{author}{\bibfnamefont{R.}~\bibnamefont{Albert}},
	\bibinfo{journal}{Science} \textbf{\bibinfo{volume}{286}},
	\bibinfo{pages}{509} (\bibinfo{year}{1999}).
	
\end{thebibliography}

\end{document}